\documentclass{article}

\usepackage{PRIMEarxiv}

\usepackage[utf8]{inputenc} 
\usepackage[T1]{fontenc}    
\usepackage{hyperref}       
\usepackage{url}            
\usepackage{booktabs}       
\usepackage{amsfonts}       
\usepackage{nicefrac}       
\usepackage{microtype}      
\usepackage{lipsum}
\usepackage{fancyhdr}       
\usepackage{graphicx}       
\usepackage{amsmath, amssymb, amsthm}
\usepackage{multirow, multicol}
\graphicspath{{media/}}     

\title{Demographic Distribution Matching between real world
and virtual phantom population
}

\author{Dhrubajyoti Ghosh, Fakrul Islam Tushar, Lavsen Dahal,
Liesbeth Vancoillie, \\ \And Kyle J. Lafata, Ehsan Samei, Joseph Y. Lo,
Sheng Luo}

\begin{document}
\maketitle

\begin{abstract}
\textbf{Purpose:} Virtual imaging trials (VITs) offer scalable and cost-effective tools for evaluating imaging systems and protocols. However, their translational impact depends on rigorous comparability between virtual and real-world populations. This study introduces DISTINCT (Distributional Subsampling for Covariate-Targeted Alignment), a statistical framework for selecting demographically aligned subsamples from large clinical datasets to support robust comparisons with virtual cohorts.

\textbf{Methods:} We applied DISTINCT to the National Lung Screening Trial (NLST) and a companion virtual trial dataset (VLST). The algorithm jointly aligned typical continuous (age, BMI) and categorical (sex, race, ethnicity) variables by constructing multidimensional bins based on discretized covariates. For a given target size, DISTINCT samples individuals to match the joint demographic distribution of the reference population. We evaluated the demographic similarity between VLST and progressively larger NLST subsamples using Wasserstein and Kolmogorov–Smirnov (K–S) distances to identify the maximal subsample size with acceptable alignment.

\textbf{Results:} The algorithm identified a maximal aligned NLST subsample of 9,974 participants, preserving demographic similarity to the VLST population. Receiver operating characteristic (ROC) analysis using risk scores for lung cancer detection showed that area under the curve (AUC) estimates stabilized beyond 6,000 participants, confirming the sufficiency of aligned subsamples for virtual imaging trial evaluation. Stratified AUC analysis revealed substantial performance variation across demographic subgroups, reinforcing the importance of covariate alignment in comparative studies.

\textbf{Conclusion:} DISTINCT provides a statistically rigorous and scalable approach for covariate alignment between real and virtual imaging cohorts based on demographic factors of variability. Although demonstrated for lung cancer screening with low-dose CT, the framework is broadly applicable to other imaging modalities and diseases, and across wide ranges of factors of variability. By enabling fair and representative performance assessments, DISTINCT advances the integration of VITs into imaging research and protocol optimization workflows.
\end{abstract}


\section{Introduction}
Lung cancer remains a leading cause of cancer mortality worldwide, underscoring the critical need for effective early detection and screening strategies \cite{ferlay2021cancer,sung2021global,siegel2023cancer}. Low-dose computed tomography (LDCT) has significantly improved early detection, motivated by the National Lung Screening Trial (NLST), which demonstrated reduced mortality among high-risk individuals \cite{national2011reduced}. However, it has been suggested that the demographic homogeneity of the NLST cohort, predominantly white participants, limits the generalizability of its findings across more diverse populations \cite{tammemagi2013selection,national2019lung}. Demographic biases in terms of anatomical variability associated with age, sex, and body mass index (BMI) can influence diagnostic performance. Failure to account for these differences can lead to biased image-based models, dose factors, and clinical recommendations, ultimately exacerbating disparities in underrepresented patient populations.

To complement traditional clinical trials, virtual imaging trials (VITs) have emerged as powerful tools. By using computationally generated human phantoms\cite{kainz2018advances, abadi2018modeling, dahal2025xcat}, VITs enable efficient simulation of patient anatomy, disease progression, and imaging protocols \cite{badano2018evaluation,abadi2020virtual,tushar2023data}. These trials circumvent many logistical, financial, and ethical constraints associated with large-scale human studies. The Virtual Lung Screening Trial (VLST) exemplifies this approach, leveraging high-fidelity phantoms to systematically evaluate LDCT performance, protocol optimization, and image quality under controlled conditions \cite{tushar2024virtual,tushar2025virtual,islam2024vlst}. However, phantom populations can differ demographically from real-world cohorts such as the NLST, potentially introducing confounding factors. Demographic variables like age and BMI affect radiation dose deposition, lung nodule detectability, and reconstruction accuracy, which are core concerns in image fidelity. Therefore, aligning the demographic composition of virtual and clinical populations is essential for unbiased and generalizable conclusions.

Addressing these discrepancies requires statistical approaches capable of capturing full distributional differences rather than focusing solely on central tendencies. Conventional methods such as t-tests or Wilcoxon tests assess mean or median differences but overlook distributional variation. In contrast, more comprehensive metrics, such as the Kolmogorov-Smirnov (K-S) and Wasserstein distances, quantify differences between entire distributions. The K-S distance measures the maximum divergence between empirical cumulative distribution functions \cite{rachev1991probability}, while the Wasserstein distance, rooted in optimal transport theory, quantifies the total “effort” required to morph one distribution into another \cite{ramdas2017wasserstein,villani2009optimal}. These measures offer more detailed insights when assessing population-level comparability in imaging trials.

Currently, there is no standardized methodological framework for ensuring demographic comparability between virtual phantoms and clinical cohorts. While several methods exist for aligning two clinical cohorts, such as propensity score matching, inverse probability weighting, and entropy balancing, these primarily target mean balance or low-order moments and fail to align the full joint distribution of covariates. This gap limits the translational relevance of VIT findings. To address this challenge, we introduce DISTINCT (Distributional Subsampling for Covariate-Targeted Alignment), a statistical algorithm that selects demographically aligned subsamples from large clinical datasets to match other populations, including virtual populations. DISTINCT uses both Wasserstein and K-S distances to quantify and minimize covariate discrepancies, enabling fair and statistically rigorous comparisons between imaging populations. This alignment improves the external validity of VIT results, supporting reliable assessments of imaging protocols, radiation dose strategies, and diagnostic performance across diverse populations. While we focus on LDCT for lung cancer screening, DISTINCT is broadly applicable to other variabilities, imaging modalities, clinical domains and certain factors of variability, facilitating equitable and representative virtual trials.

The remainder of this article is structured as follows: Section~\ref{sec:methods} details our methodology, including Section~\ref{sec:compVar} on demographic alignment metrics, Section~\ref{sec:distinct} on the DISTINCT algorithm, and Section~\ref{sec:vct} on demographic discrepancies between VLST and NLST. Section~\ref{sec:results} presents results of applying DISTINCT to align NLST with VLST. Section~\ref{sec:disc} discusses implications for imaging physics and outlines future directions, and Section~\ref{sec:conclusion} concludes the article.

\section{Methods}
\label{sec:methods}
To ensure a rigorous and quantitative assessment of demographic differences between the NLST and the VLST populations, we employed advanced statistical techniques to measure distributional divergence and developed a novel algorithm to facilitate demographically aligned dataset comparisons. This section presents the statistical metrics used to assess variability, outlines the DISTINCT algorithm for generating matched subsamples, and describes the procedure for aligning virtual and clinical trial populations.

\subsection{Statistical Metrics for Assessing Distributional Differences}
\label{sec:compVar}

In medical physics research, accurately comparing demographic distributions requires statistical tools that assess more than just central tendency. Common approaches such as the t-test or Wilcoxon rank-sum test compare means or medians, but often fail to detect broader differences in shape, skewness, or modality, features that directly impact image quality, radiation dose, and model generalizability. To capture these broader patterns, we use two complementary distributional distance metrics: the Kolmogorov–Smirnov (K-S) distance and the Wasserstein distance.

The K-S distance is a non-parametric measure that quantifies the maximum absolute difference between the empirical cumulative distribution functions (ECDFs) of two samples. For two empirical distributions \( F_A(x) \) and \( F_B(x) \), the K-S distance is defined as
\[
d_{KS}(A, B) = \sup_{x} \left| F_A(x) - F_B(x) \right|.
\]
To assess statistical significance, the associated $p$-value is approximated by
\[
P(d_{KS}(A, B) > t) = Q_{KS} \left( \sqrt{ \frac{n_A n_B}{n_A + n_B} } \cdot t \right),
\]
where \( Q_{KS} \) is the complementary cumulative distribution function of the Kolmogorov distribution, and \( n_A \), \( n_B \) denote the sample sizes. The K-S test is particularly sensitive to differences in both location and shape.

The Wasserstein distance, also known as the Earth Mover’s Distance, is derived from optimal transport theory. It measures the minimal cost required to transform one probability distribution into another by shifting probability mass across the sample space. The 1-Wasserstein distance for univariate distributions is given by
\[
d_W(A, B) = \int_0^1 \left| F_A^{-1}(p) - F_B^{-1}(p) \right| dp,
\]
where \( F^{-1} \) represents the quantile function. Unlike the K-S metric, the Wasserstein distance accounts for the magnitude of differences throughout the entire distribution, providing a more comprehensive summary of dissimilarity.

To evaluate the significance of the observed Wasserstein distance, we use a permutation-based method. This involves pooling the two datasets, randomly shuffling the group labels, and computing Wasserstein distances under the null hypothesis that the samples are from the same distribution. The empirical $p$-value is computed as
\[
P(d_W > t) = \frac{1 + \#\{ d_{W, j}^{\pi} > t \}}{1 + m},
\]
where \( d_{W, j}^{\pi} \) denotes the Wasserstein distance computed under the \( j^{\text{th}} \) permutation and \( m \) is the number of permutations performed.

These two statistical metrics provide robust and interpretable assessments of demographic divergence between the NLST and VLST populations. They form the foundation for the matching strategy described in the following section.

\subsection{Demographic Matching via the DISTINCT Algorithm}
\label{sec:distinct}

To reduce confounding from demographic imbalance between study populations, we introduce the DISTINCT (Distribution Integrator for Statistical Consistency and Normativity Technique) algorithm. DISTINCT constructs a subsample from a larger dataset that matches the demographic composition of a smaller target population. It accommodates both continuous and discrete demographic variables by discretizing continuous features and combining all variables into a unified binning structure for stratified sampling.

Let $v_C$ represent the values for a set $\mathcal{C}$ of continuous variables (e.g., age, BMI), and let $v_D$ denote values for a set $\mathcal{D}$ of discrete variables (e.g., sex, race, ethnicity). Each continuous variable $v_c \in v_C$ is discretized into $g$ bins using a binning function $f_g(v_c)$, yielding integer-valued labels. The resulting discretized vector is $f_g(v_C) = \{f_g(v_c) \mid c \in \mathcal{C}\}$. We then concatenate $v_D$ and $f_g(v_C)$ to form a unified demographic label for each individual.

As an example, a subject who is Female (label: 0), Non-Hispanic (label: 1), Asian (label: 3), with age 62 and BMI 27, would receive the demographic label $(0, 1, 3, 2, 3)$ if age and BMI are binned into intervals (55--60, 60--65, 65--70, 70--75) and (10--18.5, 18.5--25, 25--30, 30+), respectively.

\begin{figure}[h!]
    \centering
    \includegraphics[width=0.8\linewidth]{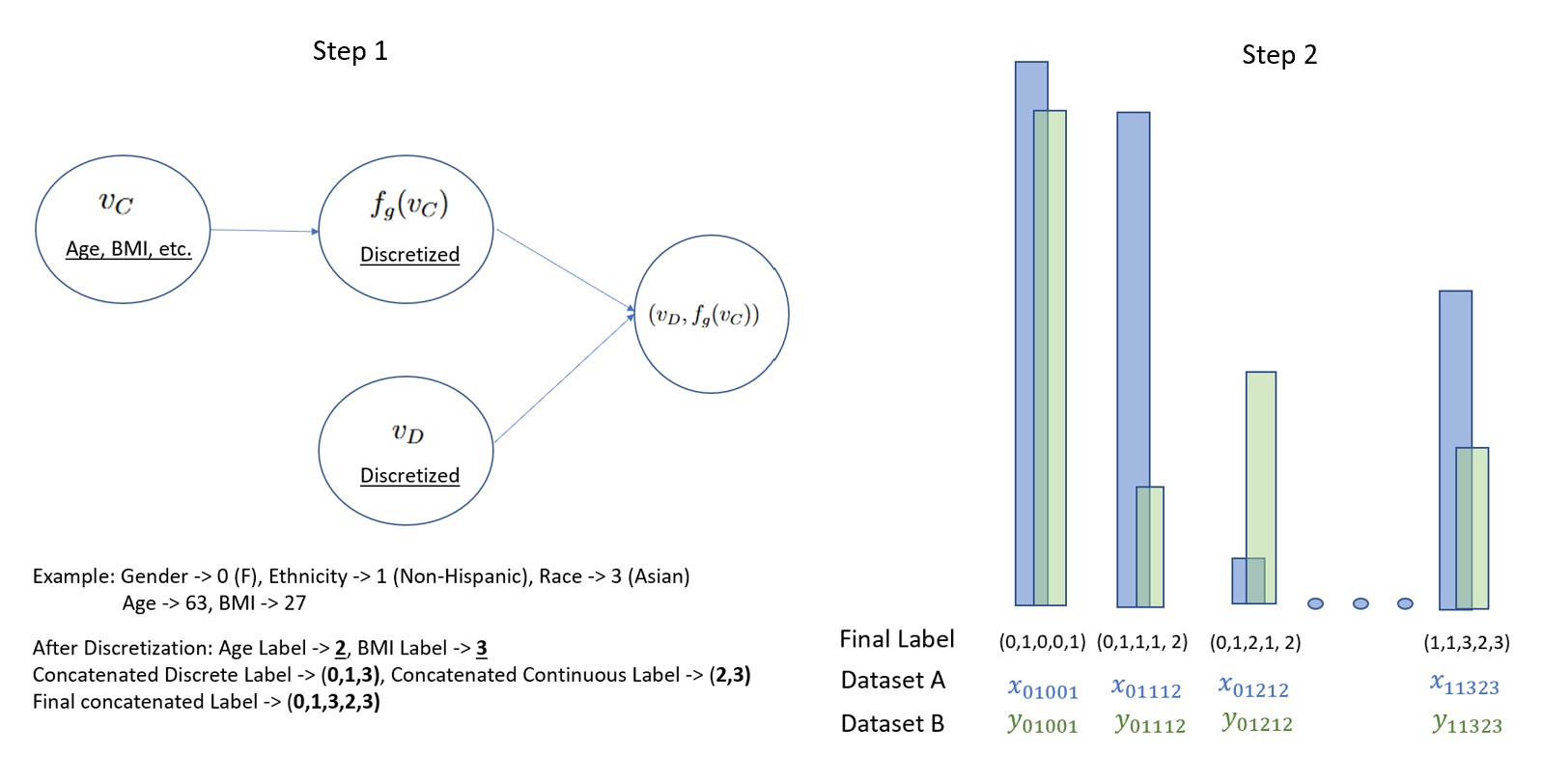}
    \caption{Schematic overview of the DISTINCT algorithm. Demographic variables, both continuous (e.g., age, BMI) and discrete (e.g., sex, race, ethnicity), are discretized and combined to define stratification bins. The algorithm constructs a subsample from the larger dataset to match the demographic bin distribution of a smaller target dataset. \newline}
    \label{fig:label-algo}
\end{figure}

The DISTINCT algorithm proceeds as follows:

\begin{enumerate}
    \item \textbf{Define demographic bins.}  
    Combine discretized continuous and categorical variables into unified bin labels. Each unique label defines a demographic stratum.

    \item \textbf{Compute target bin proportions.}  
    For each bin $l$, compute $p_l = y_l / N_T$, where $y_l$ is the number of individuals in bin $l$ in the target dataset, and $N_T$ is the total size of the target dataset.

    \item \textbf{Sample from the larger population.}  
    For a given proposed subsample size $N$, calculate the target count per bin as $\lfloor N p_l \rfloor$. If the larger dataset contains at least that many individuals in bin $l$ (i.e., $x_l \geq \lfloor N p_l \rfloor$), sample $\lfloor N p_l \rfloor$ individuals at random; otherwise, include all $x_l$ individuals.

    \item \textbf{Construct the subsample.}  
    Aggregate sampled individuals across all bins to form the final subsample. Note that due to bin constraints, the realized sample size may differ slightly from $N$.

    \item \textbf{Assess demographic alignment.}  
    Use the Wasserstein or Kolmogorov–Smirnov (K-S) test to assess whether the subsample is statistically comparable to the target population in terms of demographic distribution.

    \item \textbf{Iterate to maximize alignment.}  
    If alignment is achieved, increase $N$ and repeat the procedure to identify the largest possible aligned subset. If alignment fails, reduce $N$ and repeat until a maximally sized demographically matched subset is identified.
\end{enumerate}

As a concrete example, suppose the demographic feature space includes three discrete variables (two binary and one with five levels), and two continuous variables, each binned into five intervals (labeled 0-4). This yields $2^2 \times 5^3 = 500$ distinct demographic bins. Let $x_l$ denote the number of individuals in bin $l$ in the larger dataset, and $y_l$ the corresponding count in the target dataset. With target dataset size $N_T$, we compute $p_l = y_l / N_T$. For any proposed subsample size $N$, we aim to include $\lfloor N p_l \rfloor$ individuals per bin. If $x_l \geq \lfloor N p_l \rfloor$, we sample $\lfloor N p_l \rfloor$ individuals randomly; otherwise, we include all $x_l$.

This bin-wise sampling approach generates a subsample that closely mirrors the demographic distribution of the target dataset. Statistical tests verify alignment, and the algorithm iteratively adjusts $N$ to identify the largest demographically compatible subset.

\subsection{Demographic Comparison and Subsampling for Virtual Clinical Trials}
\label{sec:vct}

\begin{figure}[htbp]
    \centering
    \includegraphics[width=\textwidth]{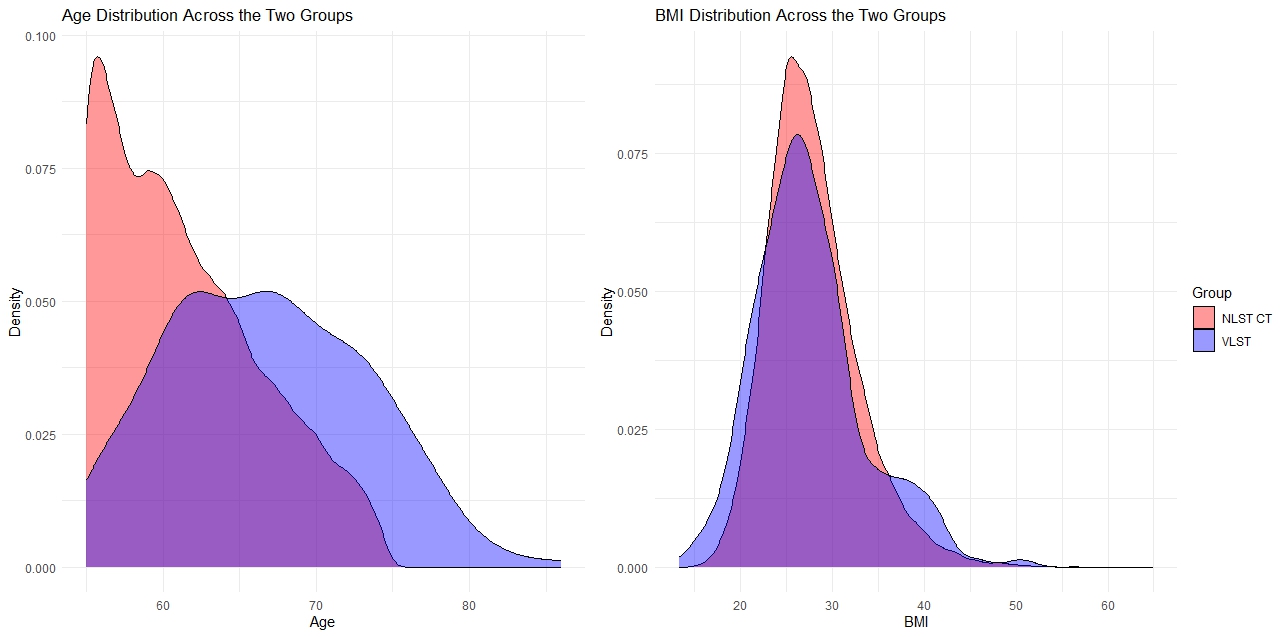}
    \caption{Histograms of age (lower truncated at 55) and BMI for NLST CT and VLST cohorts, showing significant distributional differences between the two populations. \newline}
    \label{fig:histo-cont}
\end{figure}

Tushar et al.~\cite{tushar2025virtual} introduced the Virtual Lung Screening Trial (VLST), which employs computationally representative virtual human phantoms derived from clinical CT/PET scans collected at Duke University Medical Center, a multi-hospital academic health system. The VLST dataset includes $264$ virtual patients with a mean age of $59.53$ years, $55.68\%$ male, and $98.5\%$ identifying as non-Hispanic. The average BMI is $27.13$. Baseline demographic characteristics are summarized in Table~\ref{tab:demo_nlst_vlst}, while distributions of age and BMI in both the VLST and NLST cohorts are shown in Figure~\ref{fig:histo-cont}.

Statistical comparisons between the two populations using the Wasserstein test reveal marked distributional differences. Specifically, the $p$-values for age, race, and BMI are $< 10^{-4}$, $< 10^{-3}$, and $< 10^{-3}$, respectively, while ethnicity and sex show less pronounced discrepancies ($p = 0.103$ and $0.283$, respectively). These results highlight substantial demographic mismatches that may confound downstream imaging and modeling analyses if not properly addressed.

\vskip 5 pt

\begin{table}[h!]
\centering
\begin{tabular}{lcc}
\toprule
& \textbf{NLST (LDCT, $N = 26{,}722$)} & \textbf{VLST ($N = 264$)} \\
\midrule
\textbf{Age (years)} & 61.42 $\pm$ 5.03 (43–75) & 59.53 $\pm$ 14.44 (2–86) \\
\textbf{Female – no. (\%)} & 10,953 (40.98) & 117 (44.32) \\
\textbf{Race – no. (\%)} & \begin{tabular}[t]{@{}l@{}}White: 24,289 (90.9)\\ Black: 1,195 (4.5)\\ Asian: 559 (2.1)\\ Other/Unknown: 687 (2.5)\end{tabular} & \begin{tabular}[t]{@{}l@{}}White: 198 (75.0)\\ Black: 56 (21.2)\\ Asian: –\\ Other/Unknown: 10 (3.8)\end{tabular} \\
\textbf{Non-Hispanic – no. (\%)} & 26,079 (97.4) & 260 (98.5) \\
\bottomrule
\end{tabular}
\caption{Baseline demographic characteristics of the NLST and VLST cohorts.\newline}
\label{tab:demo_nlst_vlst}
\end{table}

To enable meaningful statistical comparisons and eliminate confounding, we use the DISTINCT algorithm to extract demographically aligned subsamples from the NLST dataset that match the VLST distribution. Given that the real-world NLST cohort ($N = 26{,}722$) is much larger than the VLST cohort ($N = 264$), subsampling from NLST is computationally \newline \vskip 2 pt 
feasible and appropriate in the current setting. While it would be ideal to sample from an expanded virtual cohort, current computational limits constrain the size of VLST. However, with ongoing phantom development, the VLST is expected to eventually exceed the size of most clinical datasets, at which point demographic alignment in the reverse direction will be feasible. \newline

The DISTINCT framework is designed to operate flexibly regardless of which population is larger, making it suitable for both current and future demographic alignment tasks. In this analysis, we apply DISTINCT to extract NLST subsamples that closely mirror the VLST demographic profile, thereby enabling statistically valid and demographically balanced comparisons of imaging protocols and modeling outcomes.

\section{Results}\label{sec:results}

We applied the DISTINCT algorithm to generate demographically aligned subsamples from the NLST dataset that closely match the demographic structure of the VLST population. Given the substantial difference in sample sizes between the two cohorts (NLST: 26{,}722 participants in the LDCT arm; VLST: 264 virtual phantoms), we extracted a series of progressively larger subsamples from the NLST cohort. At each subsample size, we evaluated demographic similarity using formal statistical tests. This iterative procedure enabled us to determine the largest NLST subsample that exhibited no statistically significant differences from the VLST population across key demographic variables.

The demographic alignment process accounted for continuous variables (age and BMI) and categorical variables (sex, race, and ethnicity). To ensure computational tractability and consistent binning across cohorts, age was grouped into 5-year intervals starting from 55 years, while BMI was categorized into standard clinical ranges: underweight ($<18.5$), normal weight ($18.5 – 24.9$), overweight ($25.0 – 29.9)$, and obese ($\geq 30$). Categorical variables retained their native levels, including binary (e.g., sex) and multi-level (e.g., race) categories.

The DISTINCT algorithm defined multi-dimensional demographic bins by combining discretized continuous variables (age and BMI) with categorical variables (sex, race, ethnicity). For each bin $\ell$, the proportion of individuals in VLST was used to define the target sampling proportion, $p_\ell = y_\ell / N_T$, where $y_\ell$ denotes the number of individuals in bin $\ell$ and $N_T = 264$ is the total VLST cohort size. Given a proposed subsample size $N$, the algorithm selected approximately $\lfloor N p_\ell \rfloor$ individuals from each bin. If the number of available individuals $x_\ell$ in NLST exceeded the target, a random sample of size $\lfloor N p_\ell \rfloor$ was drawn; otherwise, all $x_\ell$ individuals were included. This bin-wise procedure was repeated across increasing values of $N$ to identify the maximal NLST subsample size for which demographic similarity could be maintained.

To assess demographic similarity, we applied both the Wasserstein distance and the Kolmogorov–Smirnov (K-S) test to each variable independently. Table~\ref{tab:merged-pvalues} reports the resulting $p$-values for each metric at various subsample sizes. Alignment was considered acceptable if all $p$-values exceeded 0.05, indicating no significant distributional differences. This criterion was satisfied up to a subsample size of approximately 9,974, beyond which age and BMI began to diverge significantly. At smaller sizes (e.g., 279 and 559), the algorithm achieved excellent alignment across all demographic variables, as reflected by non-significant $p$-values. These findings are expected: smaller subsamples offer greater flexibility in matching the target distribution, even in the presence of baseline population-level differences.

As subsample size increased, maintaining distributional similarity became more difficult due to inherent differences between the NLST and VLST populations. Sex alignment was maintained across all sample sizes, reflecting the similar female proportions in VLST (44.3\%) and NLST (41\%). In contrast, BMI emerged as the most limiting factor. For example, in the Wasserstein analysis, BMI alignment was rejected at $N = 11{,}963$ ($p < 0.001$). Race and ethnicity also showed decreasing alignment, though less sharply. The K-S test exhibited similar trends, with BMI again being the first variable to exceed the significance threshold. Taking all variables into account, the largest subsample that maintained demographic alignment across all five dimensions was approximately 9,974 individuals.

\begin{table}[h!]
    \centering
    \begin{tabular}{c c ccccc}
        \toprule
        \textbf{Test} & \textbf{Subsample Size} & \textbf{Age} & \textbf{Sex} & \textbf{Race} & \textbf{Ethnicity} & \textbf{BMI} \\
        \midrule
        \multirow{12}{*}{Wasserstein} 
        & 279   & 0.983  & 0.757  & 0.725  & 0.676  & 0.904 \\
        & 559   & 0.610  & 0.968  & 0.854  & 0.834  & 0.871 \\
        & 1038  & 0.344  & 0.563  & 0.895  & 0.413  & 0.388  \\
        & 2019  & 0.960  & 0.856  & 0.750  & 0.663  & 0.437  \\
        & 3998  & 0.688  & 0.744  & 0.990  & 0.977  & 0.786  \\
        & 5981  & 0.837  & 0.747  & 0.783  & 0.644  & 0.772  \\
        & 7981  & 0.259  & 0.909  & 0.919  & 0.738  & 0.193  \\
        & 9974  & 0.533  & 0.604  & 0.368  & 0.303  & 0.256  \\
        & 11963 & 0.579  & 0.494  & 0.442  & 0.210  & $<1e\text{-}3$  \\
        & 13963 & 0.213  & 0.452  & 0.058  & 0.114  & $<1e\text{-}3$  \\
        & 15965 & 0.074  & 0.176  & $<1e\text{-}3$  & 0.102  & $<1e\text{-}3$  \\
        & 17958 & 0.021  & 0.143  & $<1e\text{-}3$  & 0.084  & $<1e\text{-}3$  \\
        \midrule
        \multirow{11}{*}{K-S}
        & 279 & 0.933 & 0.951 & 0.952 & 1.000 & 0.462  \\
        & 559 & 0.993 & 0.746 & 0.979 & 0.442 & 0.296 \\
        & 1038 & 0.877 & 0.308 & 0.831 & 0.312 & 0.821 \\
        & 2019 & 0.775 & 0.924 & 0.614 & 0.205 & 0.380 \\
        & 3998 & 0.664 & 0.642 & 0.619 & 0.518 & 0.134 \\
        & 5981 & 0.749 & 0.714 & 0.757 & 0.619 & 0.101 \\
        & 7981 & 0.538 & 0.514 & 0.968 & 0.152 & 0.062 \\
        & 9974 & 0.233 & 0.706 & 0.053 & 0.314 & 0.052 \\
        & 11963 & 0.310 & 0.124 & $<1e\text{-}3$ & 0.218 & 0.011 \\
        & 13963 & 0.058 & 0.258 & $<1e\text{-}3$ & 0.198 & $<1e\text{-}3$ \\
        & 15965 & $<1e\text{-}3$ & 0.312 & $<1e\text{-}3$ & 0.252 & $<1e\text{-}3$ \\
        \bottomrule
    \end{tabular}
    \caption{P-values for Wasserstein and K-S tests comparing demographic distributions between VLST (fixed at $N=264$) and progressively larger subsamples drawn from NLST. Alignment is considered adequate if all $p$-values exceed 0.05. \newline}
    \label{tab:merged-pvalues}
\end{table}

\begin{figure}[h!]
    \centering
    \includegraphics[width=0.9\linewidth]{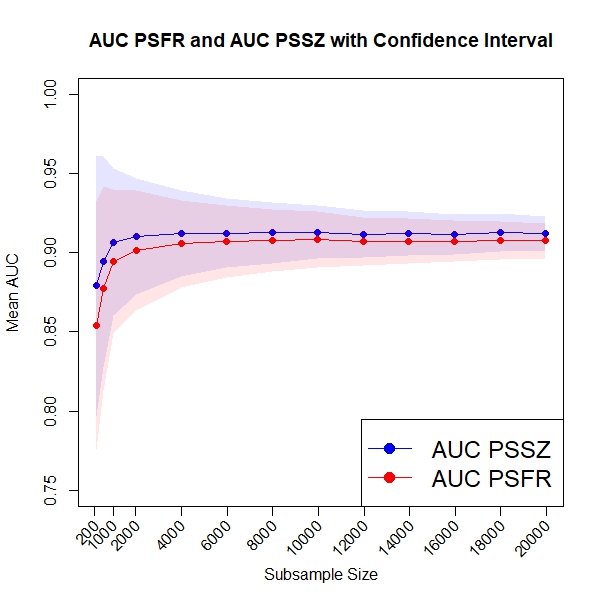}
    \caption{AUC trajectories for PSFR and PSSZ across subsamples of increasing size. Points denote mean AUCs, and shaded bands indicate 95\% confidence intervals. AUC values stabilize after approximately 6{,}000 subjects. \newline}
    \label{fig:auc1}
\end{figure}

Following subsample generation, we assessed the impact of demographic matching and subsample size on predictive model performance. We focused on two established risk scores from the NLST study developed by Pinsky et al.~\cite{pinsky2013roc}: the radiologist recommendation-based score (PSFR) and the nodule size-based score (PSSZ). PSFR captures clinical judgment based on radiologists’ follow-up recommendations, while PSSZ is an objective score derived solely from the maximum diameter of the largest non-calcified nodule. For each NLST subsample, we computed the area under the receiver operating characteristic curve (AUC) for both scores using lung cancer status as the binary outcome. Figure~\ref{fig:auc1} displays the AUC trajectories across subsample sizes. AUC estimates were highly variable at small sizes due to sampling noise, but they stabilized beyond approximately 6{,}000 individuals. PSFR plateaued around 0.91, while PSSZ stabilized near 0.92. These results suggest that demographically aligned subsamples of moderate size suffice for robust model evaluation and that gains from larger cohorts may be marginal.

To further examine subgroup-level model performance, we analyzed discrimination across demographic strata using the full NLST dataset. As detailed in Table~\ref{tab:auc} in the Appendix, females consistently showed higher AUCs than males for both PSFR (0.922 vs. 0.896) and PSSZ (0.934 vs. 0.895). Non-Hispanic participants also outperformed Hispanic individuals, with statistically significant differences in both scores. Lower BMI was associated with higher discrimination. These subgroup trends reinforce the importance of demographic alignment: without appropriate matching, apparent differences in model performance may stem from differences in population composition rather than true disparities in model quality.

In summary, the DISTINCT algorithm enables the extraction of demographically aligned subsets from large real-world clinical trials, facilitating valid comparisons with virtual trial populations. The stability of model performance across matched subsets underscores the algorithm's utility for cross-cohort analysis and highlights the potential of virtual cohorts in translational cancer research.

\section{Discussion} \label{sec:disc}

This study presents DISTINCT, a data-driven algorithm for subsampling a source population to closely match the joint distribution of covariates in a target population. DISTINCT offers a model-agnostic and nonparametric framework for demographic alignment. Unlike regression-based methods that rely on strong parametric assumptions or require labeled outcomes in both datasets, DISTINCT uses only covariate information from the target cohort to guide sample selection in the source cohort. It leverages multidimensional histogram-based binning, adaptive importance reweighting, and iterative subset construction to identify the largest feasible sample whose covariate distribution approximates that of the target population.

Empirical results using real-world clinical data from lung cancer screening demonstrate that DISTINCT generates demographically aligned subsamples that preserve downstream performance metrics, such as the area under the ROC curve (AUC) for established risk prediction models. Unlike standard matching methods, which often focus on matching mean or median values or rely solely on pairwise matching, DISTINCT aligns the full joint distribution of continuous and categorical covariates using a combination of binning and distributional distance metrics. This approach ensures better global distributional alignment and avoids common issues of residual imbalance in higher moments or joint structures. Furthermore, the iterative design of DISTINCT identifies the largest subsample size that maintains statistical alignment with the target population, offering a more interpretable and scalable alternative to conventional matching or reweighting techniques, particularly when exact matching is infeasible in high-dimensional settings.

The motivation for this approach arose from the need to enable robust comparison of virtual and real-world imaging data, where discrepancies in population composition can confound estimates of diagnostic performance. In virtual imaging trials, source data are often synthetically generated and may not reflect the diversity of real patient populations. This mismatch poses challenges for generalizability and fairness. By enabling demographic alignment, DISTINCT allows researchers to conduct trials under comparable population structures. More broadly, the method can be applied in a wide range of settings, including algorithm fairness auditing, synthetic data validation, external control arm construction, and domain adaptation in transfer learning.

Our current formulation of DISTINCT assumes that all variables used for binning and subsampling are observed in both the source and target populations. In practice, missingness in covariates or differences in measurement platforms may necessitate imputation or dimension reduction strategies. While DISTINCT can be applied using a reduced subset of harmonized covariates, this may affect its ability to align more complex joint distributions including variability factors beyond demographics. Additionally, the choice of bin granularity and overlap thresholds affects both performance and sample size, suggesting a need for adaptive or data-driven tuning procedures. Future work may explore extensions of DISTINCT for dynamic datasets, continuous recruitment, or longitudinal matching.

Together, these contributions position DISTINCT as a flexible tool for aligning populations across disparate datasets in a way that preserves statistical comparability and enhances downstream inference. Its ability to operate without requiring outcome labels in the target population makes it particularly valuable in real-world scenarios where only covariate-level data are available. By providing a principled approach to demographic alignment, DISTINCT strengthens the foundation for fair and valid comparisons in translational and clinical research.

\section{Conclusion}
\label{sec:conclusion}

The DISTINCT algorithm offers a statistically rigorous and scalable solution for aligning covariates—particularly demographic characteristics—between virtual imaging trials and real-world clinical datasets. By enabling the construction of demographically matched subsamples, DISTINCT supports robust, unbiased comparisons of predictive models and imaging performance metrics across heterogeneous populations. Our results demonstrate that modestly sized, well-matched subsets can yield stable evaluation metrics, reducing the need for unnecessarily large cohorts. Beyond demographics, DISTINCT can be readily adapted to align other clinically or physically relevant covariates, enhancing its utility in protocol optimization, phantom generation, and validation of AI-assisted imaging tools. As virtual trials become more prevalent in translational research and regulatory science, algorithms such as DISTINCT will be instrumental in ensuring methodological validity and clinical relevance.

\section*{Acknowledgments}

The study was supported by a grant from the National Institutes of Health NIBIB P41 EB028744.

\bibliographystyle{unsrt}  
\bibliography{references}  

\newpage
\appendix
\section{Appendix}
\subsection{Baseline comparability of LDCT and CXR arms in NLST}

The National Lung Screening Trial (NLST) was a landmark clinical study designed to evaluate the effectiveness of low-dose helical computed tomography (LDCT) compared to standard chest radiography (CXR) in reducing lung cancer mortality. The trial enrolled over 53{,}000 high-risk individuals, primarily heavy smokers, who were randomized to receive three annual screenings with either LDCT or CXR. The results of the NLST have had a lasting impact on lung cancer screening guidelines and early detection strategies.

Table~\ref{tab:demo_char_nlst} presents the baseline demographic characteristics of participants in the LDCT and CXR arms. To evaluate the comparability of the two groups, we assessed the distribution of key demographic variables, including age, sex, race, ethnicity, and body mass index (BMI). Statistical comparisons using the Wasserstein and Kolmogorov–Smirnov (K-S) tests revealed no significant differences between the two cohorts, with all $p$-values exceeding 0.05. This demographic equivalence supports valid cross-arm comparisons of outcomes such as diagnostic accuracy and ROC curves by reducing the likelihood of confounding from population structure.

\begin{table}[htbp]
\centering
\renewcommand{\arraystretch}{1.3}
\begin{tabular}{@{}p{4.8cm}cc@{}}
\toprule
\textbf{Characteristic} & 
\begin{tabular}[c]{@{}c@{}}LDCT Arm\\ (N = 26{,}722)\end{tabular} & 
\begin{tabular}[c]{@{}c@{}}CXR Arm\\ (N = 26{,}730)\end{tabular} \\
\midrule
\begin{tabular}[c]{@{}l@{}}Age (years)\\ Mean $\pm$ SD (Min--Max)\end{tabular} & 
61.42 $\pm$ 5.03 (43--75) & 
61.42 $\pm$ 5.02 (49--79) \\
\midrule
\begin{tabular}[c]{@{}l@{}}Sex\\ Female, no. (\%)\end{tabular} & 
10{,}953 (40.98\%) & 
10{,}969 (41.05\%) \\
\midrule
\begin{tabular}[c]{@{}l@{}}Race\\ White, no. (\%)\\ Black/African American, no. (\%)\\ Asian, no. (\%)\\ Other/Unknown, no. (\%)\end{tabular} & 
\begin{tabular}[c]{@{}c@{}}24{,}289 (90.9\%)\\ 1{,}195 (4.5\%)\\ 559 (2.1\%)\\ 687 (2.6\%)\end{tabular} & 
\begin{tabular}[c]{@{}c@{}}24{,}260 (90.8\%)\\ 1{,}181 (4.4\%)\\ 536 (2.0\%)\\ 745 (2.8\%)\end{tabular} \\
\midrule
\begin{tabular}[c]{@{}l@{}}Ethnicity\\ Non-Hispanic, no. (\%)\end{tabular} & 
26{,}079 (97.4\%) & 
26{,}039 (97.6\%) \\
\bottomrule
\end{tabular}
\caption{Baseline demographic characteristics of participants in the LDCT and CXR arms of the NLST. Statistical tests indicated no significant differences between the two groups across all variables assessed.}
\label{tab:demo_char_nlst}
\end{table}

\subsection{Stratified AUC analysis of NLST risk scores by demographic subgroup}

The National Lung Screening Trial (NLST) demonstrated that low-dose computed tomography (LDCT) screening reduced lung cancer mortality by 20\% compared to chest radiography (CXR) among individuals aged 55 to 74 years \cite{national2011reduced}. Although approximately 60\% of NLST participants were male, the trial demonstrated comparable mortality reductions for both sexes \cite{national2019lung}. The study population was predominantly white, raising concerns about generalizability to underrepresented racial and ethnic groups \cite{tammemagi2013selection}.

To examine how predictive performance varies across demographic subgroups, we analyzed two lung cancer risk scores introduced by Pinsky et al. \cite{pinsky2013roc}: the radiologist recommendation–based propensity score (PSFR) and the size–based propensity score (PSSZ). PSFR reflects radiologists’ follow-up recommendations and incorporates subjective imaging features such as shape and density. In contrast, PSSZ is computed solely from the diameter of the largest non-calcified nodule (NCN), providing an objective and standardized risk measure.

Pinsky et al. reported that PSFR achieved a slightly higher area under the receiver operating characteristic curve (AUC = 0.934) than PSSZ (AUC = 0.928), suggesting that radiologist judgment added incremental diagnostic value. Adjusting score thresholds improved specificity with limited sensitivity trade-offs. For example, a PSFR threshold of 3+ yielded 92.4\% specificity and 86.9\% sensitivity, while a PSSZ threshold of 8 mm resulted in 92.0\% specificity and 83.2\% sensitivity.

Table~\ref{tab:auc} reports AUC values stratified by demographic subgroup using the NLST Spiral CT data. Females exhibited higher AUCs than males for both PSFR (0.922 vs.\ 0.896) and PSSZ (0.934 vs.\ 0.895), with $p < 10^{-5}$ based on Delong’s test. Similarly, non-Hispanic participants outperformed Hispanic participants (PSFR: 0.931 vs.\ 0.905; PSSZ: 0.965 vs.\ 0.909; $p < 0.001$). These findings underscore the influence of demographic composition on risk model performance and reinforce the value of demographic alignment. In our study, the DISTINCT algorithm was applied to construct demographically matched NLST subsamples, supporting unbiased cross-comparisons between real and virtual imaging datasets.

\begin{table}[htbp]
\centering
\begin{tabular}{ccccc}
\toprule
& & & \multicolumn{2}{c}{AUC} \\
\cline{4-5}
\multicolumn{3}{c}{Demographics (N = 26,722)} & PSFR & PSSZ \\
\midrule
\multirow{4}{*}{Age} & \multicolumn{2}{c}{55--60 (11,440)} & 0.909 $\pm$ 0.005 & 0.912 $\pm$ 0.005 \\
                     & \multicolumn{2}{c}{60--65 (8,170)} & 0.903 $\pm$ 0.005 & 0.908 $\pm$ 0.005 \\
                     & \multicolumn{2}{c}{65--70 (4,756)} & 0.901 $\pm$ 0.007 & 0.902 $\pm$ 0.007 \\
                     & \multicolumn{2}{c}{70--75 (2,353)} & 0.911 $\pm$ 0.009 & 0.914 $\pm$ 0.009 \\
\midrule
\multirow{2}{*}{Sex} & \multicolumn{2}{c}{Male (15,769)} & 0.896 $\pm$ 0.004 & 0.895 $\pm$ 0.005 \\
                     & \multicolumn{2}{c}{Female (10,953)} & 0.922 $\pm$ 0.004 & 0.934 $\pm$ 0.004 \\
\midrule
\multirow{2}{*}{Ethnicity} & \multicolumn{2}{c}{Non-Hispanic (25,788)} & 0.931 $\pm$ 0.019 & 0.965 $\pm$ 0.012 \\
                           & \multicolumn{2}{c}{Hispanic (445)} & 0.905 $\pm$ 0.003 & 0.909 $\pm$ 0.003 \\
\midrule
\multirow{4}{*}{BMI} & \multicolumn{2}{c}{10--18.5 (227)} & 0.901 $\pm$ 0.016 & 0.894 $\pm$ 0.012 \\
                     & \multicolumn{2}{c}{18.5--25 (7,434)} & 0.920 $\pm$ 0.005 & 0.926 $\pm$ 0.005 \\
                     & \multicolumn{2}{c}{25--30 (11,143)} & 0.898 $\pm$ 0.005 & 0.900 $\pm$ 0.005 \\
                     & \multicolumn{2}{c}{$>$30 (7,434)} & 0.892 $\pm$ 0.007 & 0.900 $\pm$ 0.007 \\
\midrule
\multicolumn{3}{l}{Full Dataset} & 0.910 $\pm$ 0.005 & 0.920 $\pm$ 0.005 \\
\bottomrule
\end{tabular}
\caption{Stratified area under the ROC curve (AUC) for PSFR and PSSZ risk scores across demographic subgroups in the NLST Spiral CT arm.}
\label{tab:auc}
\end{table}

\end{document}